\documentclass[prb,aps,showpacs]{revtex4}
\usepackage{graphicx, amsmath, amsthm}
\makeatletter
\begin{document}
\title{Method of studying the Bogoliubov-de Gennes equations for the superconducting vortex lattice state}

\author{Qiang Han}
\affiliation{Department of Physics, Renmin University, Beijing,
China}

\date{\today}
\begin{abstract}
In this paper, we present a method to construct the eigenspace of the normal-state electrons moving in a 2D square lattice in presence of a perpendicular uniform magnetic field which imposes (quasi)-periodic boundary conditions for the wave functions in the magnetic unit cell. An exact unitary transformations are put forward to correlate the discrete eigenvectors of the 2D electrons with those of the Harper's equation. The cyclic-tridiagonal matrix associated with the Harper's equation is then tridiagonalized by another unitary transformation. The obtained eigenbasis is utilized to expand the Bogoliubov-de Gennes equations for the superconducting vortex lattice state, which showing the merit of our method in studying the large-sized system. To test our method, we have applied our results to study the vortex lattice state of an $s$-wave superconductor.
\end{abstract}

%\pacs{}
\maketitle

\section{Introduction}

Vortex states of type-II superconductors has received greater attention in recent years. Theoretical formalism describing this effect is the Bogoliubov-de Gennes (BdG) approach~\cite{bdg}, which can be viewed as real-space extension of the Bardeen-Cooper-Schrieffer (BCS) theory. This method allows one to reveal effects of imperfections in superconductors, such as impurities, surfaces, as well as field-induced vortices which we are concerned with in this paper.
In recent years there are numerous studies on the superconducting vortex lattice state by solving the discrete
BdG equations and consequently diagonalization of the BdG mean-field Hamiltonian on a two-dimensional tight-binding
lattice \cite{ywang,mtakigawa,qhan,mfranz,jxzhu}. However, the size of the unit cell of the vortex lattice, which is inversely proportional to the amplitude of the magnetic field, is limited by computer resources since the dimension of the BdG equations grows with system size.
Therefore early numerical works on small-size unit cells are limited to high magnetic fields over ten Tesla, which is stronger than used in most experiments, and no remarkable progress has been made over the past decade due to the time consuming of full diagonalization (i.e. all eigenvalues and eigenvectors) of the mean-field BdG Hamiltonian. In fact in BCS-type superconductors, electrons near the Fermi level bind into Cooper pairs by exchanging virtual bosons such as phonons, excitons or plasmons etc. Therefore, there exists an energy cutoff, which equals approximately the characteristic energy of the bosons such as the Debye phonon frequency of conventional superconductors, and correspondingly only the electronic states lying near the Fermi surface within an energy shell are necessary to be explored. For the vortex problem, the most appropriate starting point is to find the relevant electronic states which participate in BCS pairing and forming of the superconducting vortex lattice when an external magnetic field is applied. The eigenequation describing this states is a 2D difference equation formulated on a magnetic unit cell which is twice the size of that of the superconducting vortex lattice. This eigenvalue problem is also demanding when the system size is large even though only a truncated eigenspace is desired.

In this paper we present an exact reduction of the Hermitian matrix associated with the 2D discrete equation into a tridiagonal matrix, which composed of two consecutive unitary transformations. First we reduces the 2D discrete equation that describes electrons moving in a magnetic unit cell into the famous Harper's equation. Algebraically, this unitary transformation reduces the Hermitian matrix into a cyclic tridiagonal matrix corresponding to the Harper's equation. Then by another exact transformation the cyclic tridiagonal matrix is further reduced into a tridiagonal form. The exact reduction greatly lessens the computational burden in numerical methods. Ultimately we diagonalize the tridiagonal matrix utilizing available standard software packages to find the appropriate eigenstates near the Fermi level, i.e. the truncated eigenspace, and then expand and diagonalize the BdG equations in this truncated eigenbasis.

This paper is organized as follows. In Section \ref{sec2}, we derive the BdG equations expanded in terms of the truncated eigenbasis of the normal-state electrons in the magnetic field. The Hermitian matrix associated with
the 2D tight-binding electrons on a 2D square lattice in a magnetic field is reduced into a tridiagonal form in Section \ref{sec3}. In Section \ref{sec4}, the vortex lattice state of an $s$-wave superconductor is studied as a test of our method. Section \ref{sec5} gives the concluding remarks.

\section{The BdG equations for vortex lattice states}
\label{sec2}

In this work, we adopt a BCS-type mean-field Hamiltonian defined on a two-dimensional(2D) square lattice,
\begin{equation}
    \hat{H}=\hat{H}_0+\hat{H}_\Delta
           =\sum_\mathbf{i,j,\sigma}
                \left(t_\mathbf{ij}-\mu \delta_\mathbf{i,j}\right) c_{\mathbf{i}\sigma}^\dagger c_{\mathbf{j}\sigma}
                    +
            \sum_{\mathbf{i,j}} \left(
                \Delta_\mathbf{ij} c_{\mathbf{i}\uparrow}^\dagger  c_{\mathbf{j}\downarrow}^\dagger + \text{H.c.}
                                \right)
\end{equation}
where $\Delta_\mathbf{ij}=\frac{V}{2}\langle c_{\mathbf{i}\uparrow}c_{\mathbf{j}\downarrow}- c_{\mathbf{i}\downarrow}c_{\mathbf{j}\uparrow}\rangle $ for spin-singlet pairing\cite{triplet}. In an external uniform magnetic field applied in the $z$-direction, the hopping integral acquires the
Peierls phase factor as
\begin{equation}
    t_\mathbf{ij}=-t\exp\left(i\frac{2\pi}{\phi_0}\int_\mathbf{j}^\mathbf{i}\mathbf{A}\cdot d\mathbf{l}\right)
            =\left\{
                \begin{array}{ll}
                    -t, &  \mathbf{i}=(m_x,m_y), \ \ \ \mathbf{j}=(m_x+1,m_y) \\
                    -t\exp\left(i\frac{2\pi Ba^2}{\phi_0} m_x \right), & \mathbf{i}=(m_x,m_y), \ \ \ \mathbf{j}=(m_x,m_y+1)
                \end{array}
            \right.
            .
\end{equation}
Here $t$ denotes the nearest-neighbor hopping integral. We choose the Landau gause with $\mathbf{A}=B(0,x,0)$ and the screening field induced by the supercurrent is neglected for extreme type-II
superconductors. $\phi_0=h/e$ is the electronic flux quantum.
Hereafter we use pair of integers $\mathbf{i}\equiv(m_x,m_y)$ as index of the site in the square lattice to denote the $x$ and $y$ coordinates.
In the Nambu representation, the above Hamiltonian can be written as
\begin{eqnarray}
    \hat{H}
        &=&\sum_{\mathbf{i,j}}
                \left(
                    c_{\mathbf{i}\uparrow}^\dagger, c_{\mathbf{i}\downarrow}
                \right)
                \left(
                    \begin{array}{cc}
                        t_\mathbf{ij}-\mu\delta_\mathbf{ij} & \Delta_\mathbf{ij} \\
                        \Delta_\mathbf{ij}^{*} & -t_\mathbf{ij}^{*}+\mu\delta_\mathbf{ij}
                    \end{array}
                \right)
                \left(
                    \begin{array}{c}
                        c_{\mathbf{j}\uparrow} \\
                        c_{\mathbf{j}\downarrow}^\dagger
                    \end{array}
                \right) \\
        &=&
            \hat{\Psi}^\dagger
                \left(
                    \begin{array}{cc}
                        \check{h}-\mu\check{I} & \check{\Delta} \\
                        \check{\Delta}^* & -\check{h}^*+\mu\check{I}
                    \end{array}
                \right)
            \hat{\Psi}
\end{eqnarray}
where $\hat{\Psi}^\dagger(\hat{\Psi})$ is the Nambu creation (annihilation) operator defined as
$\hat{\Psi}^\dagger=(c_{\mathbf{1}\uparrow}^\dagger,c_{\mathbf{1}\downarrow}, \cdots, c_{\mathbf{i}\uparrow}^\dagger,c_{\mathbf{i}\downarrow}, \cdots, c_{K\uparrow}^\dagger,c_{K\downarrow})$ with $K$ the total number of lattice sites.
$\check{h}$ and $\check{\Delta}$ are $K\times K$ matrices with elements
$(\check{h})_\mathbf{ij}=t_\mathbf{ij}$ and $(\check{\Delta})_\mathbf{ij}=\Delta_\mathbf{ij}$, respectively. $\check{I}$ is the $K\times K$ identity matrix.
The mean-field Hamiltonian can be diagonalized by solving the following BdG equations,
\begin{equation}
    \sum_{\mathbf{j}}
        \left(
            \begin{array}{cc}
                t_\mathbf{ij}-\mu\delta_\mathbf{ij} & \Delta_{\mathbf{ij}} \\
                \Delta_{\mathbf{ij}}^{*} & -t_\mathbf{ij}^{*}+\mu\delta_\mathbf{ij}
            \end{array}
        \right)
        \left(
            \begin{array}{c}
                u_n(\mathbf{j}) \\
                v_n(\mathbf{j})
            \end{array}
        \right)
    =E_n
        \left(
            \begin{array}{c}
                u_n(\mathbf{i}) \\
                v_n(\mathbf{i})
            \end{array}
        \right)
        \label{bdgexpanded}
\end{equation}
which can be viewed as Schrodinger-like equations for the electron and hole amplitudes of a BdG quasiparticle. The pairing
potential $\Delta_\mathbf{ij}$ couples the $u$ and $v$ components and satisfies the self-consistent condition
\begin{equation}
    \Delta_\mathbf{ij}=V\sum_{|E_n|<E_D} u_n(\mathbf{i}) v_n^*(\mathbf{j}) \tanh\left(\frac{E_n}{2k_\text{B}T}\right),
\end{equation}
where $E_D$ is the Debye-type cutoff energy of the pairing interaction. The BdG equations (\ref{bdgexpanded}) can
be expressed compactly in a matrix form
\begin{equation}
    \left(
        \begin{array}{cc}
            \check{h}-\mu\check{I} & \check{\Delta} \\
            \check{\Delta}^* & -\check{h}^*+\mu\check{I}
        \end{array}
    \right)
    \left(
        \begin{array}{c}
            u \\
            v
        \end{array}
    \right)
    = E
    \left(
        \begin{array}{c}
            u \\
            v
        \end{array}
    \right).
    \label{bdgcompact}
\end{equation}
with $u$ and $v$ $K$-dimensional vectors.

Abrikosov vortices, each of which carries one superconducting flux quantum $\Phi_0=h/2e$, are created and form a lattice structure in a type-II superconductor if one applies a magnetic field ($B_{c1} \leq B \leq B_{c2} $). The vortex lattice causes periodic modulation of the pairing potential and accordingly yields energy bands of BdG quaiparticles. To study this effect in our study we adopt the concept of magnetic unit cell (MUC) whose size is twice that of the unit cell of the vortex lattice and accordingly each MUC accommodates one electronic flux quantum $\phi_0=2\Phi_0$. Here for illumination of our method, we study the square vortex lattice which is aligned with the underlying crystalline lattice. The unit cell size of the vortex lattice is $N_x \times N_x$ , corresponding to a uniform magnetic field $B=\Phi_0/(N_x a)^2$. Each MUC accommodates two adjacent vortices in the $y$ direction. Therefore the MUC is of size $N_x \times N_y$ with $N_y = 2 N_x$. The whole system is composed of $M_x \times M_y$ MUC's. Thus the whole system has $M_x M_y N_x N_y$ lattice sites.  For later convenience, we introduce
a dimensionless parameter $\alpha\equiv Ba^2/\phi_0=1/(N_x N_y)$ denoting
the ratio of magnetic flux per plaquette to the electronic flux quantum $\phi_0$. 
%And the range of the pair of integers $(m_x,m_y)$ of the lattice site in the MUC are $m_x=0,1,\ldots N_x-1$ and $m_y=0,1,\ldots N_y-1$.

In the Abrikosov vortex lattice state, the BdG equations (\ref{bdgexpanded}) is symmetric under magnetic translation with the translation vector
$\mathbf{R}= l_x N_x\mathbf{e}_x + l_y N_y\mathbf{e}_y$. Due to this magnetic translational symmetry in the $x$ and $y$ direction, the quasiparticle amplitudes can be expressed in the magnetic Bloch form as
\begin{equation}
\left(
\begin{array}{c}
u(\mathbf{i}) \\
v(\mathbf{i})
\end{array}
\right)=
e^{i\mathbf{k}\cdot\mathbf{i}}
\left(
\begin{array}{c}
 u^\mathbf{k}(\mathbf{i}) \\
 v^\mathbf{k}(\mathbf{i})
\end{array}
\right)
\end{equation}
where the magnetic Bloch wave vector $\mathbf{k}=\frac{2\pi l_x}{M_x N_x}\mathbf{e}_x+\frac{2\pi l_y}{M_y N_y}\mathbf{e}_y$ with $l_{x,y}=0,1,\cdots,M_{x,y}-1$.
%%$\mathbf{k}$ lies in the first Brillouin zone
This transformation reduces Eq.~(\ref{bdgcompact}) to the new BdG equations for $u^\mathbf{k}$ and $v^\mathbf{k}$
\begin{equation}
        \left[
        \begin{array}{cc}
            \check{h}^\mathbf{k}-\mu\check{I} & \check{\Delta}^\mathbf{k} \\
            (\check{\Delta}^{-\mathbf{k}})^* & -(\check{h}^{-\mathbf{k}})^*+\mu\check{I}
        \end{array}
    \right]
    \left(
        \begin{array}{c}
            u_n^\mathbf{k} \\
            v_n^\mathbf{k}
        \end{array}
    \right)
    =E_n^\mathbf{k}
    \left(
        \begin{array}{c}
            u_n^\mathbf{k} \\
            v_n^\mathbf{k}
        \end{array}
    \right)
\label{newbdg}
\end{equation}
where the matrix elements of the $\mathbf{k}$-dependent matrices $\check{h}^\mathbf{k}$ and $\check{\Delta}^\mathbf{k}$ are
$
(\check{h}^\mathbf{k})_\mathbf{ij} = t_\mathbf{ij}e^{-i\mathbf{k}\cdot(\mathbf{i-j})},
(\check{\Delta}^{\mathbf{k}})_\mathbf{ij} = \Delta_\mathbf{ij} e^{-i\mathbf{k}\cdot(\mathbf{i-j})},
$
The quasiparticle amplitudes $u^\mathbf{k}$ and
$v^\mathbf{k}$ satisfy the quasi-periodic boundary conditions with period $N_x$ along the $x$ direction
\begin{eqnarray}
    u^\mathbf{k}(m_x + N_x, m_y) &=& e^{-i2\pi m_y N_x \alpha} u^\mathbf{k}(m_x, m_y) \label{bu} \\
    v^\mathbf{k}(m_x + N_x, m_y) &=& e^{i2\pi m_y N_x \alpha} v^\mathbf{k}(m_x, m_y)  \label{bv}
\end{eqnarray}
while they are periodic in the $y$ direction with period $N_y$. The $(m_x,m_y)$'s in Eqs.~(\ref{newbdg},\ref{bu},\ref{bv}) are restricted to sites within one MUC with $m_{x,y}=0,1,\cdots,N_{x,y}-1$. The above
procedure reduces the Hermitian matrix with linear dimension $2 M_x M_y N_x N_y$ [ Eq.~(\ref{bdgcompact}) ] into  direct sum of $M_x M_y$ block matrices, each of which is labeled by $\mathbf{k}$ and has linear dimension $2 N_x N_y$ [ Eq.~(\ref{newbdg}) ]. For each quasimomentum $\mathbf{k}$, Eq.~(\ref{newbdg})
is diagonalized along with the boundary conditions and then the whole solutions of all $\mathbf{k}$ are used by the following
equation
\begin{equation}
    \Delta_\mathbf{ij}=V\sum_{|E_n^\mathbf{k}|<E_D} u_n^\mathbf{k}(\mathbf{i})    [v_n^{\mathbf{k}}(\mathbf{j})]^* e^{i\mathbf{k}\cdot(\mathbf{i-j})} \tanh\left(\frac{E_n^\mathbf{k}}{2k_\text{B}T}\right),
    \label{selfc}
\end{equation}
to achieve self-consistence.

In the literature typical size of the unit cell of the vortex lattice studied by previous works was limited around $20\times 20$ \cite{ywang,mtakigawa,qhan,mfranz,jxzhu}. Such a small unit cell size corresponds to a magnetic field as large as $B=\Phi_0/(20a)^2\approx 32$ Tesla, which is much higher than used by most experiments, if one assumes a typical lattice constant $a\approx$4{\AA}.
%no remarkable progress was made in the last decade due to the time consuming of full diagonalization (i.e. all %eigenvalues and eigenvectors) of the BdG Hamiltonian (Eq.~\ref{newbdg}). 
Therefore one should find the way to diagonalize the BdG Hamiltonian (Hermitian matrix) with larger scale in order to match numerical calculation with experimental data. Although one can take advantage of the sparse nature of the BdG Hamiltonian $\check{\Omega}$, we think that iterative methods, such as the Lanczos algorithm, are not appropriate for this
problem because they are designed to compute a few eigenvalues(eigenvectors) with largest/smallest magnitudes.

To study the vortex lattice state with larger unit cell and correspondingly weaker and realistic magnetic field, we re-express the real-space BdG equations Eq.~(\ref{newbdg}) in the diagonal representation of $\check{h}^\mathbf{k}$ which describes the 2D tight-binding electrons in presence of a magnetic field,
\begin{equation}
    \check{h}^\mathbf{k}\varphi_q^\mathbf{k} = \varepsilon_q^\mathbf{k} \varphi_q^\mathbf{k},
\label{2DHarper}
\end{equation}
where $\varphi_q^\mathbf{k}$ obeys the same boundary condition as Eq.(\ref{bu}).
According to the BCS theory, only a fraction of electrons in the energy shell $E_D$ around the Fermi energy participate in the Cooper pairing.
Therefore we should first get the eigenstates $\varphi_q^\mathbf{k}$ from Eq.~(\ref{2DHarper}) with energies
$|\varepsilon_q^\mathbf{k}-\mu|\leq E_D$ relative to the Fermi level, which will be addressed in Section \ref{sec3}.
Here the quasiparticle amplitudes
$u^\mathbf{k}$ and $v^\mathbf{k}$ are expanded in the basis functions $\varphi_q^\mathbf{k}$ and $(\varphi_q^{-\mathbf{k}})^*$, respectively,
\begin{equation}
    \left\{
        \begin{array}{l}
            u_n^\mathbf{k} = \sum_q a_{n}^\mathbf{k}(q) \varphi_q^\mathbf{k} \\
            v_n^\mathbf{k} = \sum_q b_{n}^\mathbf{k}(q) (\varphi_q^{\mathbf{-k}})^*
        \end{array}
    \right. .
    \label{expansion}
\end{equation}
This reduces Eq.~(\ref{newbdg}) to,
\begin{equation}
    \sum_q
    \left[
        \begin{array}{cc}
            (\varepsilon_q^\mathbf{k}-\mu) \delta_{p,q}  &  \Delta_{p,q}^\mathbf{k} \\
            \Delta_{p,q}^{-\mathbf{k}*} &  (\mu-\varepsilon_q^\mathbf{k}) \delta_{p,q}
        \end{array}
    \right]
    \left[
        \begin{array}{c}
            a_{n}^\mathbf{k}(q) \\
            b_{n}^\mathbf{k}(q)
        \end{array}
    \right]
    = E_n^\mathbf{k}
    \left[
           \begin{array}{c}
            a_{n}^\mathbf{k}(p) \\
            b_{n}^\mathbf{k}(p)
        \end{array}
    \right]
\label{bdgespace}
\end{equation}
where the matrix element $\Delta_{p,q}^\mathbf{k}$ is calculated according to,
\begin{equation}
    \Delta_{p,q}^\mathbf{k} = (\varphi_p^\mathbf{k})^\dagger \check{\Delta}^\mathbf{k} (\varphi_q^{-\mathbf{k}})^* = \sum_\mathbf{i,j} [\varphi_p^\mathbf{k}(\mathbf{i})]^* \Delta_\mathbf{ij} e^{-i\mathbf{k}\cdot(\mathbf{i-j})} [\varphi_q^{-\mathbf{k}}(\mathbf{j})]^*,
    \label{me}
\end{equation}
while from Eqs.~(\ref{selfc}) and (\ref{expansion}), we have
\begin{equation}
    \Delta_\mathbf{ij} = V \sum_{\mathbf{k},p,q,n} \varphi_p^\mathbf{k}(\mathbf{i}) \varphi_q^\mathbf{-k}(\mathbf{j}) a_{n}^\mathbf{k}(p) [b_{n}^{\mathbf{k}}(q)]^*
        \tanh \left( \frac{E_n^\mathbf{k}}{2k_\text{B} T} \right)
    \label{selfc1}
\end{equation}
The Eqs.~(\ref{bdgespace})-(\ref{selfc1}) are solved iteratively until self consistence is satisfied. Eventually we can calculated the local density of states, which is proportional to the differential tunneling conductance, from the energy spectrum and wave functions,
\begin{equation}
    \rho(\mathbf{i},E) = \sum_{\mathbf{k},n} |u_n^\mathbf{k}(\mathbf{i})|^2\delta(E-E_n^\mathbf{k})+|v_n^\mathbf{k}(\mathbf{i})|^2\delta(E+E_n^\mathbf{k}).
\end{equation}
At the present stage, we have expressed the BdG equations in the truncated eigenbasis of $\check{h}^\mathbf{k}$. The issue now is how to compute this truncated eigenbasis, i.e. the eigenstates of $\check{h}^\mathbf{k}$ lying within an energy shell $E_D$ around the Fermi level.

Utilizing standard computational algorithm~\cite{numrecipe}, it will be rather time-consuming to compute some selected eigenstates  of a large matrix as $\check{h}^\mathbf{k}$, whose size $N \times N$ grows fastly with the length scale of the MUC, by tridiagonalizing the matrix {\it numerically}.  Even after taking advantage of the sparse nature of $\check{h}^\mathbf{k}$, we find that the iterative methods, such as the Lanczos algorithm, are not quite appropriate for this problem because they are most efficient for finding largest/smallest eigenvalues(eigenvectors). In the following sections, we solve this issue by showing that $\check{h}^\mathbf{k}$ can be tridiagonalized {\it exactly} by two unitary transformations. Then we appeal to standard packages such as LAPACK~\cite{lapack} to compute the desired eigenstates of the resulting tridiagonal matrix within an energy range.

\section{2D tight-binding electrons in a magnetic unit cell}
\label{sec3}

In this section, we show in detail the method of reduce the matrix $\check{h}^\mathbf{k}$ to a tridiagonal matrix exactly. The eigenequation of $\varphi_n^\mathbf{k}$ [Eq.~(\ref{2DHarper})], i.e. the discrete Schro\"{o}dinger equation describing a 2D free electron moving in a perpendicular uniform magnetic field in a square lattice, can be written in an explicit form
\begin{equation}
    e^{ik_x} \varphi_n^\mathbf{k}(m_x+1,m_y) + e^{-ik_x} \varphi_n^\mathbf{k}(m_x-1,m_y)
    +e^{i(2\pi m_x \alpha + k_y)} \varphi_n^\mathbf{k}(m_x,m_y+1) + e^{-i(2\pi m_x \alpha + k_y)} \varphi_n^\mathbf{k}(m_x,m_y-1) =  \varphi_n^\mathbf{k}(m_x,m_y),
\label{2DmagneticBloch}
\end{equation}
where $\tilde{\varepsilon}_n^\mathbf{k}=\varepsilon_n^\mathbf{k}/(-t)$ and $\varphi^\mathbf{k}$ obeys the quasi-periodic boundary condition along the $x$ direction and periodic boundary condition along the $y$ direction
\begin{equation}
    \left\{
        \begin{array}{l}
            \varphi^\mathbf{k}(m_x+N_x, m_y) =  e^{-i2\pi m_y N_x \alpha} \varphi^\mathbf{k}(m_x,m_y), \\
            \varphi^\mathbf{k}(m_x, m_y+N_y) =  \varphi^\mathbf{k}(m_x,m_y).
        \end{array}
    \right.
    \label{2Dboundary}
\end{equation}
First we find that the eigenfunction $\varphi^\mathbf{k}$ is related to the eigenfunction $g^\mathbf{k}$ of
the Harper's equation by a unitary transformation. Explicitly,
\begin{equation}
    \varphi_n^\mathbf{k}(m_x,m_y) = \frac{1}{\sqrt{N_y}}\sum_{l=0}^{N_y-1} e^{i 2\pi m_y lN_x \alpha} g_n^\mathbf{k}(m_x + l N_x).
    \label{trans1}
\end{equation}
Substituting the above equation into Eq.~(\ref{2DmagneticBloch}), one readily find that $g_n^\mathbf{k}$ satisfies the Harper equation
\begin{equation}
    e^{ik_x} g_n^\mathbf{k}(m+1) + e^{-ik_x} g_n^\mathbf{k}(m-1) + 2\cos(2\pi m\alpha + k_y) g_n^\mathbf{k}(m) = \tilde{\varepsilon}_n^\mathbf{k} g_n^\mathbf{k}(m),
\label{eigen-g}
\end{equation}
Here $m=0,1,\cdots,N-1$ with $N=N_xN_y$. $g$ satisfies the periodic boundary condition $g_n^\mathbf{k}({m+N})=g_n^\mathbf{k}(m)$. In the matrix form, the Harper equation can be expressed as
\begin{equation}
    \check{P}^\mathbf{k}g_n^\mathbf{k}=\tilde{\varepsilon}_n^\mathbf{k} g_n^\mathbf{k},
\end{equation}
where
\begin{equation}
    \check{P}^\mathbf{k}=\left(
        \begin{array}{cccccc}
          a_0 & e^{ik_x} &   &   &   & e^{-ik_x} \\
          e^{-ik_x} & a_1 & e^{ik_x} &   &   &   \\
            & e^{-ik_x} & \cdot & \cdot &   &   \\
            &   & \cdot & \cdot & \cdot &   \\
            &   &   & \cdot & \cdot & e^{ik_x} \\
          e^{ik_x} &   &   &   & e^{-ik_x} & a_{N-1}
        \end{array}
    \right),
\end{equation}
with $a_m=2\cos(2\pi m\alpha+k_y)$. And the eigenvector $g_n^\mathbf{k}=(g_n^\mathbf{k}(0),\cdots,g_n^\mathbf{k}(N-1))^\text{T}$.
The Harper's equation can be viewed as discrete analogue of the
Schr\"{o}dinger equation of the one-dimensional quantum harmonic oscillator.
Therefore, the energy eigenfunctions $g_n^\mathbf{k}$ are discrete analogues of the Hermit-Gaussian-type wave functions.

The periodic tridiagonal matrix $\check{P}^\mathbf{k}$ can be further reduced to a tridiagonal matrix by another unitary transformation.
For simplicity, we only show the procedure  for the $k_y=0$ case and the following discussion can be readily generalized for $k_y \neq 0$.
The transformation of wave vectors from $g$ to $f$ is as follows,
\begin{equation}
 \left\{
 \begin{array}{l}
  f(0) = g(0), \\
  f(1) = \frac{e^{ik_x}g(1) + e^{-ik_x} g(N-1)}{\sqrt{2}}, \\
  f(2) = \frac{e^{2ik_x}g(2) + e^{-2ik_x} g(N-2)}{\sqrt{2}}, \\
  \cdots  \\
  f(\frac{N}{2}-1) = \frac{e^{i(\frac{N}{2}-1)k_x} g(\frac{N}{2}-1) + e^{-i(\frac{N}{2}-1)k_x} g(\frac{N}{2}+1)}{\sqrt{2}}, \\
  f(\frac{N}{2}) = g(\frac{N}{2}), \\
  f(\frac{N}{2}+1) = \frac{e^{i(N/2-1)k_x} g(\frac{N}{2}-1) - e^{-i(\frac{N}{2}-1)k_x} g(\frac{N}{2}+1)}{\sqrt{2}}, \\
  \cdots \\
  f(N-2) = \frac{e^{2ik_x}g(2) - e^{-2ik_x} g(N-2)}{\sqrt{2}}, \\
  f(N-1) = \frac{e^{ik_x}g(1) - e^{-ik_x} g(N-1)}{\sqrt{2}}.
 \end{array}
 \right.
 \label{trans2}
\end{equation}
Substituting the above relations into Eq.~(\ref{eigen-g}), we have the eigenequation for $f$, which reads,
\begin{equation}
    \check{T}^\mathbf{k} f_n^\mathbf{k} = \tilde{\varepsilon}_n^\mathbf{k} f_n^\mathbf{k}
\end{equation}
where $\check{T}$ is an $N\times N$ tridiagonal matrix,
\begin{equation}
    \check{T}^\mathbf{k} =
    \left(
         \begin{array}{ccccccccccccc}
              a_0 & \sqrt{2} & & & & & & & & & & & \\
              \sqrt{2} & a_1 & 1 & & & & & & & & & & \\
               & 1 & \cdot & \cdot & & & & & & & & & \\
               & & \cdot & \cdot & \cdot &   &  & & & & & & \\
               & & & \cdot & \cdot  & 1 & & & & & & & \\
               & & & & 1 & a_{\frac{N}{2}-1} & \sqrt{2}\cos(\frac{N}{2}k_x) & & & & & & \\
               & & & & & \sqrt{2}\cos(\frac{N}{2}k_x) & a_{\frac{N}{2}} & -i\sqrt{2}\sin(\frac{N}{2}k_x) & & & & & \\
               & & & & & & i\sqrt{2}\sin(\frac{N}{2}k_x) & a_{\frac{N}{2}-1} & 1 &  &   &  &   \\
               & & & & & & & 1 & a_{\frac{N}{2}-2} & 1 & & &  \\
               & & & & & & & & 1 & \cdot & \cdot & &   \\
               & & & & & & & & & \cdot & \cdot & \cdot & \\
               & & & & & & & & & & \cdot & \cdot & 1  \\
               & & & & & & & & & & & 1 & a_{1}
         \end{array}
    \right)
    \label{trid}
\end{equation}
and $f_n^\mathbf{k}=(f_n^\mathbf{k}(0),\cdots,f_n^\mathbf{k}(N-1))$.

After two consecutive unitary transformations, we have successfully reduce the Hermitian matrix $\check{h}^\mathbf{k}$ into a tridiagonal
matrix $\check{T}^\mathbf{k}$.
Here we emphasize that the reduction is exact without any numerical assumption and takes no CPU time compared with the numerical reduction.
Then the eigenproblem of the tridiagonal matrix $\check{T}^\mathbf{k}$ can be solved using standard packages such as LAPACK.

\section{An Example: Vortex lattice states of a type-II $s$-wave superconductor}
\label{sec4}

In this section, we illustrate how our method is applied in solving the BdG equation for the vortex lattice states of an $s$-wave superconductor. The microscopic parameters used in this paper are as follows. As a model calculation we set the relevant parameters as follows. $\mu = -3 t$ which gives rise to an almost circular Fermi surface with the Fermi wave vector $k_\text{F}\approx 1.03 a^{-1}$ and Fermi velocity $v_\text{F}\approx 1.81 ta/\hbar$. The on-site attractive interaction $V=2t$. The Debye-type energy cutoff $E_D=0.5t$. This set of parameters results in an $s$-wave pairing potential $\Delta_0\approx 0.065 t$ in the zero-temperature limit with the estimated coherence length
$\xi_0=\hbar v_\text{F}/\pi\Delta_0\approx 9 a$.

The model calculation is carried out for a system composed of $M_x\times M_y=40\times 20$ MUC's with each MUC of size $N_x\times N_y = 80\times 160$ which corresponds to a magnetic field $B=\phi_0/(N_x N_y a)^2\approx 2.0$ Tesla if the lattice constant is set as 4{\AA}. Therefore, for each $\mathbf{k}$ of the total $800$ quasimomenta, we employ standard LAPACK routine to diagonalize the $12800\times 12800$ tridiagonal matrix (Eq.~\ref{trid}) and find that there are approximately 1173 eigenstates $\{f_n^\mathbf{k}\}$ whose eigenenergies lying within the energy range $|\varepsilon^\mathbf{k}-\mu|\leq E_D$. We can obtain the eigenstates of the Harper's equation $\{g_n^\mathbf{k}\}$ by the inverse transformation of Eq.~(\ref{trans2}). Then substituting $g_n^\mathbf{k}$ into Eq.~(\ref{trans1}) we successfully obtain the truncated eigenbasis $\{\varphi_n^\mathbf{k}\}$, in which the BdG equations (\ref{bdgespace}) are expressed as the $2\times 1173$-dimensional eigenvalue problem and the matrix elements $\Delta_{p,q}^\mathbf{k}$ is calculated from Eq.~(\ref{me}). After the BdG equations are diagonalized for each $\mathbf{k}$, we substitute the quasiparticle amplitudes $a_n^\mathbf{k}$ and $b_n^\mathbf{k}$ into the self-consistent condition Eq.~(\ref{selfc1}) to compute the renewed values of the pairing potential. Eqs.~(\ref{bdgespace})-(\ref{selfc1}) are solved iteratively until convergence is reached.

\begin{figure}[ht]
    \includegraphics[width=8cm]{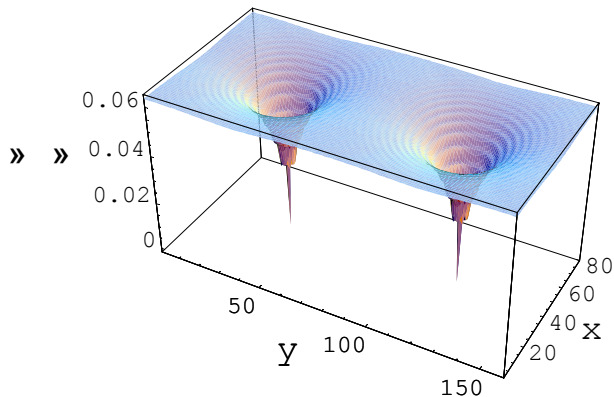}
    \caption{Spatial distribution of the magnitude of the $s$-wave pairing potential $|\Delta|$ in one magnetic unit cell of size $80\times 160$. The $x$- and $y$-axis are in unit of the lattice constant $a$. $|\Delta|$ is in unit of the hoping integral $t$.}
\label{distribution}
\end{figure}
In Fig.~\ref{distribution} we show the spatial variation of the self-consistent pair potential within one $80\times 160$-sized magnetic
unit cell, in which two superconducting vortices are situated. The $s$-wave pairing potential vanishes at the center of each of the two $80\times 80$ squares and increases with the distance from the core center and recovers to its bulk value approximately with a length scale $\xi_0$. The variation of the pairing potential around the vortex core exhibits almost circular symmetry as shown in the figure. The reasons are twofold. Firstly, the Fermi level is far away from the van Hove singularity and accordingly the Fermi surface is approximately circular. Secondly, the impact from neighboring vortices which arranged squarely is weak because the distance between the adjacent vortices is about one order of magnitude larger compared to the the characteristic coherence length. 

\begin{figure}[ht]
    \includegraphics[width=8cm]{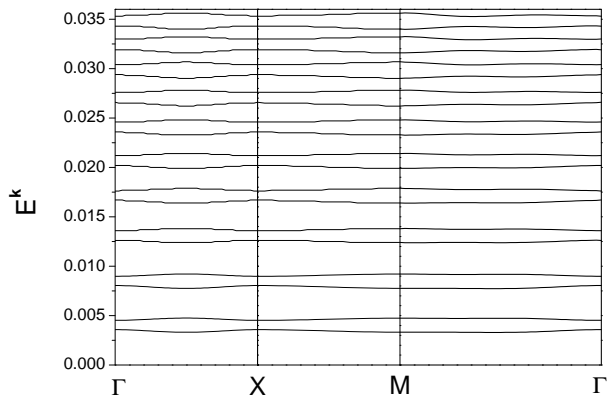}
    \caption{Quasiparticle spectrum in the magnetic Brillouin zone. See text for details.}
\label{spectrum}
\end{figure}
Figure \ref{spectrum} displays the quasiparticle spectra along three high-symmetry lines in the magnetic Brillouin zone, where $\Gamma X$, $XM$ and $M\Gamma$ connect two of the three points: $\Gamma=(0,0)$, $X=(\frac{\pi}{N_x a},0)$ and $M=(\frac{\pi}{N_x a},\frac{\pi}{N_y a})$. As shown in the figure, the vortex bound states, which is localized in a isolated vortex line as revealed in Refs.\cite{ccaroli,fgygi}, are broaden into energy bands in the superconducting vortex lattice owing to the interference effect. However due to the localized nature of the vortex states, the overlapping of the quasiparticle wave functions belonging to difference vortices is weak especially for the low-lying states. Consequently the bands with lower energies are flatter and the level spacing between pairs of the first few lowest-lying bands is of the order of $\Delta_0^2 / E_\text{F}$. These results are consistent with previous works\cite{qhan,mfranz,kyasui}.

\begin{figure}[ht]
    \includegraphics[width=8cm]{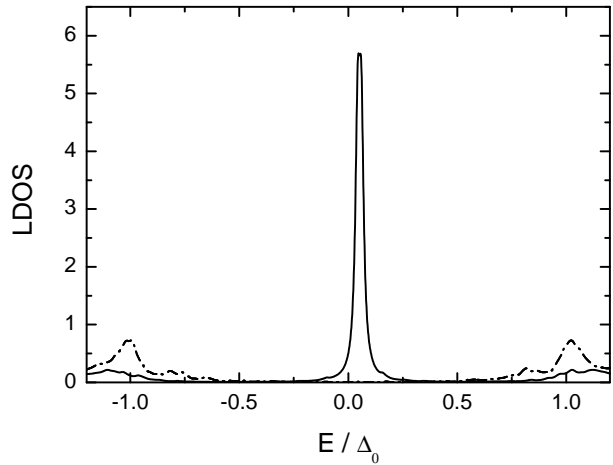}
    \caption{LDOS as a function of energy at the vortex core center (solid line) and the inter-vortex site (dash-dotted line). }
\label{figldos}
\end{figure}
In Fig.~\ref{figldos} we plot the local density of states (LDOS) as a function of energy at the vortex core center and the inter-vortex site. At the center of the vortex core, the LDOS is greatly enhanced at the energy approximately
equal to $\Delta_0^2 / E_\text{F}$ due to the strongly localized vortex bound states, while depressed around
$E=\pm\Delta_0$ as compared with the LDOS at the inter-vortex site. The model calculation shows the feasibility of our methods in studying the vortex lattice with large unit cell.

%Our results are even in good agreement with those for the isolated vortex line in the continuum limit since our model calculation are performed on relatively large magnetic unit cells and .

\section{Conclusion}
\label{sec5}

The discrete BdG equations developed in the 2D tight-binding lattice have been used to study the magnetic-field-induced superconducting vortex lattice state in the literature. The size of the system studied in previous works was limited due to the full diagonalization of the BdG hamiltonian directly. In this paper, we have extended this method by 
constructing a truncated eigenspace for the normal-state electrons moving on a 2D square lattice in presence of a uniform magnetic field. The motion of the electrons is governed by the vector potential, which impose a (quasi)-periodic boundary condition along the $x$ and $y$ directions of the magnetic unit cell. We have presented two consecutive unitary transformations to reduce the Hermitian matrix for the 2D electrons into a tridiagonal matrix exactly. By doing so, we have successfully related the desired eignbasis with that of the celebrated Harper's equation which is the eigenequation for a periodic-tridiagonal matrix. Then the second transformation is applied to further reduce the periodic tridigoanl matrix to a tridigonal one. This greatly reduces the cost of CPU time and helps us to treat systems with much larger size. To test our method and elucidate it more specifically, we have applied our results to study the vortex lattice states of an $s$-wave superconductor. The extension of our method to more sophisticated band structure as well as to
2D triangular or honeycomb lattice will be performed in future works.   

\acknowledgements
This work was supported by the NSFC grand under Grants No. 10674179.


\begin{thebibliography}{20}
\bibitem{triplet} Our method can be readily extended to the case of spin-triplet pairing.
\end{thebibliography}
\end{document}